\documentclass[aps, onecolumn,
twocolumn,
pre,superscriptaddress, 
notitlepage]{revtex4-2}
\usepackage{bm,amsmath, amssymb,graphicx, mathtools,multirow,gensymb}
\usepackage[colorlinks=true,linkcolor=MidnightBlue,
urlcolor=black,citecolor=MidnightBlue,anchorcolor=MidnightBlue]{hyperref}
\usepackage[dvipsnames]{xcolor}
\newcommand{\first}[1]{\left(#1\right)}

\usepackage{float}

\begin{document}
\title{
Boundary- and Screening-Induced Bubbly Phases in Autophoretic Active Matter
}
\author{Kingshuk Panja}
\email{ph20d205@smail.iitm.ac.in}
\affiliation{Department of Physics, Indian Institute of Technology Madras, Chennai, 600036, India}
\author{G\"{u}nther Turk}
\email{guenther.turk@princeton.edu}
\affiliation{Princeton Materials Institute, Princeton University, Princeton, NJ, 08544, USA}
\author{Rajesh Singh}
\email{rsingh@physics.iitm.ac.in}
\affiliation{Department of Physics, Indian Institute of Technology Madras, Chennai, 600036, India}
\begin{abstract}
Spatial confinement and chemical screening fundamentally reshape the non-equilibrium phase behavior of autophoretic active particles. Here, we present a systematic study mapping the collective dynamics of self-propelled particles governed by chemo-attractive translational forces ($\mu_t < 0$) and chemo-repulsive rotational torques ($\mu_r > 0$) across varying screening parameters $\kappa$, torque magnitudes $\mu_r$, and boundary condition coefficients $\Lambda^c$. Beyond standard chemotactic macro-phase separation and dynamic clustering, we report the emergence of novel boundary- and screening-induced \textit{bubbly phases}, classified into \textit{boiling} and \textit{bursting} bubbles. Using a metric triad of steady-state cluster fraction $\langle S \rangle$, temporal fluctuation magnitude $\sigma_S$, and coordination number $\langle Q \rangle$, we 
draw phase diagrams to demarcate phases for no-flux boundaries ($\Lambda^c = 1$) and chemically permeable interfaces  ($\Lambda^c = 0$). 
Increasing chemical screening ($\kappa$) systematically suppresses long-range attraction, driving sequential phase transitions from macro-scale collapse toward bubbly states, dynamic micro-clusters, and homogeneous \textit{gas} phases, while simultaneously inducing aggregate shape anisotropy. 
These findings provide predictive design rules for controlling active assembly and transport in microfluidic environments.
\end{abstract}
\maketitle

\section{Introduction}
Active matter is composed of individual units that convert energy from their surroundings into mechanical motion and thereby operate far from thermodynamic equilibrium \cite{ramaswamy2010, marchetti2013, te2026colloquium}. At the micro- and nanoscale, examples range from biological microswimmers to synthetic catalytic Janus particles. Their persistent self-propulsion, together with interactions between particles, gives rise to collective behaviours that have no direct counterpart in passive colloidal suspensions.

A central example is motility-induced phase separation (MIPS), in which active Brownian particles with purely repulsive interactions separate into dense and dilute phases \cite{cates2015motility, marchetti2016minimal, redner2013structure, fily2012athermal, bialke2013microscopic, cates2025active}. Here, phase separation requires neither attractive forces nor an aligning interaction, but arises from the feedback between persistent motion and crowding. Particle-resolved simulations of MIPS have reported dilute voids, or bubbles, within the dense phase \cite{stenhammar2014phase, patch2018, yan2025stochastic}. Related behaviour also emerges at the continuum level. In Active Model B+, non-equilibrium currents that break detailed balance can reverse the usual Ostwald process, stabilising bubbly phase separation in which dilute bubbles persist within a dense background \cite{tjhung2018cluster}. These results establish that bubbles can arise in active phase separation without phoretic interactions.

The situation is different for autophoretic particles. Surface reactions generate chemical fields that propel the particles and mediate interactions between them. Each particle therefore modifies the field experienced by its neighbours, producing long-ranged phoretic forces and torques without direct contact. Such interactions can be non-reciprocal \cite{fruchart2021non, saha2019pairing, zhang2021active} and lead to collective states beyond those of active Brownian particles with steric interactions alone. Particle-resolved and continuum studies have found chemotactic collapse, dynamic clusters, wave patterns and other forms of spatiotemporal organisation \cite{pohlStarkPRL2014, liebchen2015clustering, liebchen2017phoretic, liebchen2019interactions, saha2014clusters, saha2019pairing, singh2019competing, agudo2019active, fadda2023, subramaniam2026collective}. More recently, screened phoretic interactions have also been shown to support flocking and related collective states \cite{das2024flocking, subramaniam2025minimal, adhikary2025flocking}. In particular, phoretic attraction can produce full collapse or dynamic clustering at particle densities much lower than those typically required for MIPS \cite{pohlStarkPRL2014, liebchen2017phoretic, liebchen2019interactions}.

Despite these advances, controlled bubbly phases have, to the best of our knowledge, not been reported in particle-resolved systems with phoretic interactions. This distinction is important: the bubbles considered here are not voids generated within a purely repulsive MIPS dense phase. Instead, they arise from a competition between chemo-attractive translational interactions, chemo-repulsive rotational interactions and self-propulsion. They occur between the previously identified regimes of chemotactic collapse and dynamic clustering, and provide a microscopic realisation of bubbly active matter in a system whose interactions are mediated by a physical chemical field.

The environment provides an additional means of controlling these interactions. Chemical screening sets their effective range, while a nearby boundary modifies the chemical field through its solute boundary condition. Boundaries are known to alter the motion of individual autophoretic particles \cite{dasBoundariesCanSteer2015, uspal2015self, simmchenTopographicalPathwaysGuide2016, turk2024fluctuating,dasFloorCeilingSlidingChemically2020, turk2025autophoretic, ruangkriengsinAutophoresisJanusParticle2026a}; however, the combined effect of screening and chemical boundary conditions on the collective phase behaviour of autophoretic suspensions remains largely unexplored.

In this work, we use particle-resolved simulations to show that attractive phoretic forces and repulsive phoretic torques generate two distinct bubbly states, which we term \textit{boiling bubbles} and \textit{bursting bubbles}. By varying the screening parameter, torque strength and solute boundary condition, we map the transitions between full collapse, bubbly phases, dynamic clusters and a homogeneous gas. Comparing a no-flux surface with a chemically permeable interface shows that the boundary-mediated image field changes the phase behaviour qualitatively: a no-flux surface enlarges the range over which bubbly states occur and stabilises the boiling-bubble phase, which is absent at the permeable interface within the parameter range studied. Our results therefore demonstrate that chemical boundaries do not merely shift phase boundaries, but can select the non-equilibrium steady states available to an autophoretic suspension.

The rest of the paper is organised as follows. In Sec. \ref{sec:model}, we present the particle model, equations of motion and chemical boundary conditions. In Sec. \ref{ses:results}, we identify the observed phases and describe the mechanisms responsible for their formation. Finally, in Sec. \ref{sec:conclusion}, we summarise our main results and discuss directions for future work.
\section{Model and Method}
\label{sec:model}
\begin{figure}[t]
\centering
\includegraphics[width=0.48\textwidth]{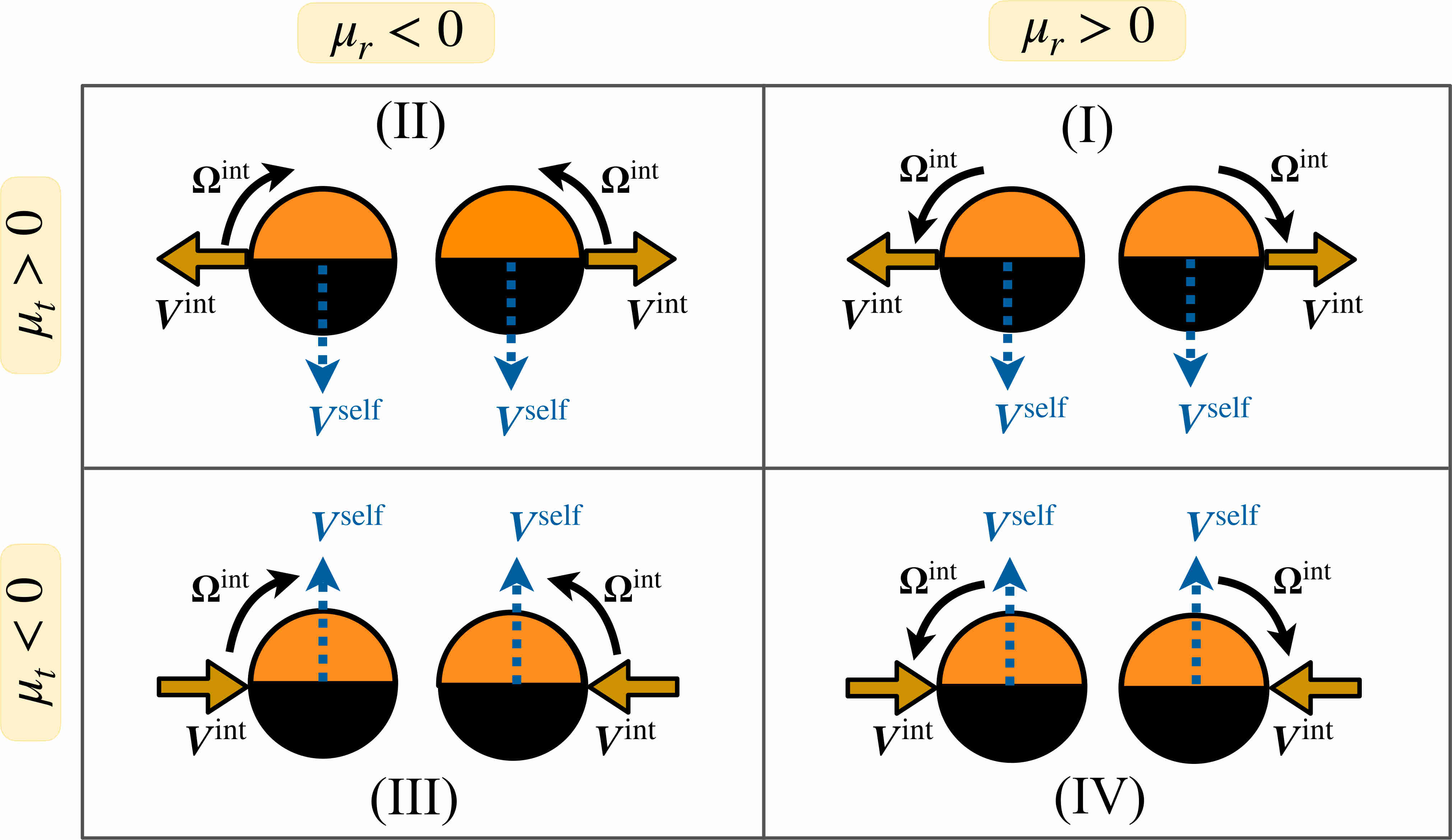}
\caption{Schematics of Janus particles, half-covered by a catalytic cap (orange), interacting with both each other and a nearby boundary.
Here, we provide a representation of the various types of phoretic interactions permitted by the model outlined in section \ref{sec:model}. In this paper, we specifically focus on region IV, where \(\mu_t < 0\) and \(\mu_r > 0\). In this region, particles are attracted to each other, but they also tend to turn away from each other (see Video - I). In this model, the direction of self-propulsion of a particle points away from the cap when \(\mu_t > 0\) and towards the cap when \(\mu_t < 0\). Furthermore, the velocity resulting from interaction with another particle is directed away from that particle for \(\mu_t > 0\) but towards the other particle for \(\mu_t < 0\). Additionally, particle caps orient themselves away from each other for \(\mu_r > 0\) and towards each other for \(\mu_r < 0\) (see Eq. \ref{eq:partE}).
}
\label{fig:schematic}
\end{figure}

We consider a system of $N$ active colloidal particles of radius $b$, each half-covered by a catalytic coating that generates a flux of solutes propelling the particle.
In Figure \ref{fig:schematic}, we show a schematic representation of the system. 
The position of the $i^\text{th}$ 
particle 
is denoted by $\bm r_{i}$, while its orientation is denoted by 
$
\bm { p }_i = \cos\theta_i \,\hat x + \sin\theta_i\,\hat y $,
pointing towards the catalytic cap.
Each particle has an intrinsic self-propulsion speed $v_0$. 
The update equations for position $\bm r_i$ and orientation $\theta_i$ of the $i$th particle are given as:
\begin{subequations}
\begin{align}
    \frac{d{\bm r}_i}{dt}&=M_t\,\bm{F}^{\mathrm p}_i+\bm V^{\mathrm{self}}_i+\bm V^{\mathrm{int}}_i,\\
    \frac{d  \theta_i}{dt}&=\Omega^{\mathrm{int}}_i+\sqrt{2D_r}\,\xi_i.
\end{align}
\label{eq:mainLE}    
\end{subequations}
Here, 
$\xi_i$ is a 
white noise with zero mean and unit variance, while $D_r$ is the strength of the noise. 
In addition, we have defined:
\begin{align}
\bm V_i^{\mathrm{self}} 
&=  
-\mu_t v_0\bm { p }_i,
\qquad
\bm V_i^{\mathrm{int}} = 
-\mu_t{2v_0}\bm J_i,
\\
\bm\Omega^{\mathrm{int}}_i
&=- \frac{9}{4}v_0\mu_r 
\left(
\bm 
{ p }_i\times \bm J_i
\right),
\label{eq:partE}
\end{align}
where $\bm \Omega_i^{\mathrm{int}} = \Omega_i^{\mathrm{int}}( \hat{\bm{ p }}_i\times \hat{\bm J}_i )$.
Here, the self-velocity $\bm V_{\mathrm{self}} $  is the same for all the particles. The interaction between the particles is given through the terms $\bm V_i^{\mathrm{int}}$ for translation and 
$\bm \Omega_i^{\mathrm{int}}$ for rotation. 
The dimensionless numbers
$\mu_t$ and $\mu_r$ are the
strengths of phoretic inter-particle forces and torques respectively. 
In Eq.\eqref{eq:partE}, 
$\bm J_i = \tfrac{|\langle\mu\rangle|}{2v_0}\bm{\nabla}c\big|_{\bm{r}_i}$ is the dimensionless chemical current, which accounts for inter-particle phoretic interactions due to the chemical field $c(\bm r, t)$. 

For a semi-infinite domain bounded by a plane surface, this current is given by (see appendix \ref{app:DeriveEOM} for a derivation):
\begin{align}
\bm J_i =  - \sum_{j=1}^N 
\Bigg[\first{\kappa + \frac{1}{r_{ij}}}\frac{e^{-\kappa r_{ij}}}{r_{ij}^2} \,\bm r_{ij}
\nonumber\\
+ \Lambda^c\first{\kappa + \frac{1}{r^*_{ij}}}\frac{e^{-\kappa r^*_{ij}}}{{r^*_{ij}}^2} 
\,\bm r^*_{ij}
\Bigg]
\label{eq:current}
\end{align}
where $\bm r_{ij}=\left(\bm r_i - \bm r_j\right)/b$ is 
the dimensionless separation between the particles
and $\bm r_{ij}^*=(\bm r_i - \bm r_j^*)/b$, 
with 
$\bm{r}_i=(x_i,y_i,h)^T$, and $\bm{r}^*_i=(x_i,y_i,-h)^T$, where the latter denotes the position of the image  \cite{turk2025autophoretic}. Here, $\kappa=b/\ell_s=b\sqrt{\lambda_d/D_1}$ is the dimensionless inverse screening length, with $\ell_s=\sqrt{D_1/\lambda_d}$ the screening length.
Such screened phoretic interactions have been recently studied in the context of flocking of autophoretic particles \cite{das2024flocking, adhikary2025flocking,adhikary2026}.
The dimensionless parameter $\Lambda^c$ is defined as:
 \begin{align}
    \Lambda^c = \frac{D_1 - \kappa_c D_2}{D_1 + \kappa_c D_2}.
\label{eq:BC}
\end{align}
The dimensionless number $\Lambda^c$  
governs the chemical boundary conditions. 
The solute permeability $\kappa_c\in\{0,\,1\}$ indicates whether the surface is impermeable ($\kappa_c=0$) or permeable ($\kappa_c=1$) to solutes.
In general, $-1 \leq \Lambda^c \leq 1$.
A no-flux wall is represented by $\Lambda^c = 1$, 
while $\Lambda^c=0$ implies a chemically permeable interface between equally diffusive media ($D_2=D_1$), i.e. the chemical field does not sense the boundary. In the latter case, equation \eqref{eq:current} simplifies to the unbounded chemical current used by \citet{pohlStarkPRL2014, liebchen2019interactions} in similar analyses of the clustering behavior of chemically interacting particles. In this paper, we consider the 
special cases: $\Lambda^c=0$, $\Lambda^c=-1$ and $\Lambda^c=1$. 
Our main focus is $\Lambda^c=1$ as it corresponds to the experimentally realisable no-flux solid boundary.
We show that the collective dynamics is sensitive to the choice of $\Lambda^c$.
The role of boundary condition was studied in Ref. \cite{dasBoundariesCanSteer2015,uspal2015self,simmchenTopographicalPathwaysGuide2016,dasFloorCeilingSlidingChemically2020,turk2025autophoretic,ruangkriengsinAutophoresisJanusParticle2026a}.
The combined effect of phoretic screening and boundary conditions to control the collective behavior of autophoretic particles 
has not been studied to the best of our knowledge.

Apart from the chemical interactions, the particle interact through a force which precludes overlap through the terms $M_t \bm F^{\mathrm p}_i$. Here, $M_t=1/(6\pi\eta b)$, with $b$ as the radius of the particle and $\eta$ as the fluid viscosity. The force $    \bm F^{\mathrm p}_{i} = -\sum_{i\neq j}\bm \nabla_i U^\text{p}(r_{ij})$ is obtained from the WCA potential \cite{weeks1971role}. Explicit form of the WCA potential 
$U^\text{WCA}$ 
is given in appendix \ref{app:simDetails}. 
This purely repulsive force ensures that particles do not overlap with each other.







\section{Results and discussion}
\label{ses:results}
In this section, we describe the main results. 
In this paper, we choose $\mu_t=-1$. In addition, we consider the case of $\mu_r>0$. Thus, we are operating in the region (IV) of the 
Figure \ref{fig:schematic}a, in which particles attract and turn away from each other. In addition, we vary boundary condition through the parameter $\Lambda^c$, which was defined in Eq.\eqref{eq:BC}. We also define the following dimensionless numbers
 \begin{align}
    \mathrm{Pe} &= v_{\scriptscriptstyle{0}} \frac{\tau}{b},
    \qquad 
    \tau = \frac{1}{D_r},
\label{eq:BC2}
\end{align}
where $\tau$ is the persistence time, and $\mathrm{Pe}$ is the P\'{e}clet number.
In this paper, we fix $\mathrm{Pe} = 100$ to study the role of screening  (through the parameter $\kappa$) of phoretic interactions and strength of phoretic torques (though the parameter $\mu_r$ defined above), while we vary the boundary conditions through the parameter $\Lambda^c$.

\subsection{Steady-state behavior}

To quantify clustering behaviour, we have used the fraction of particles in clusters, $S$, defined as:

\begin{equation}
    S = \frac{\sum_n nN_n}{N}
    \label{eq:psi}
\end{equation}

\begin{figure}
    \centering
 \includegraphics[width=0.484\textwidth]{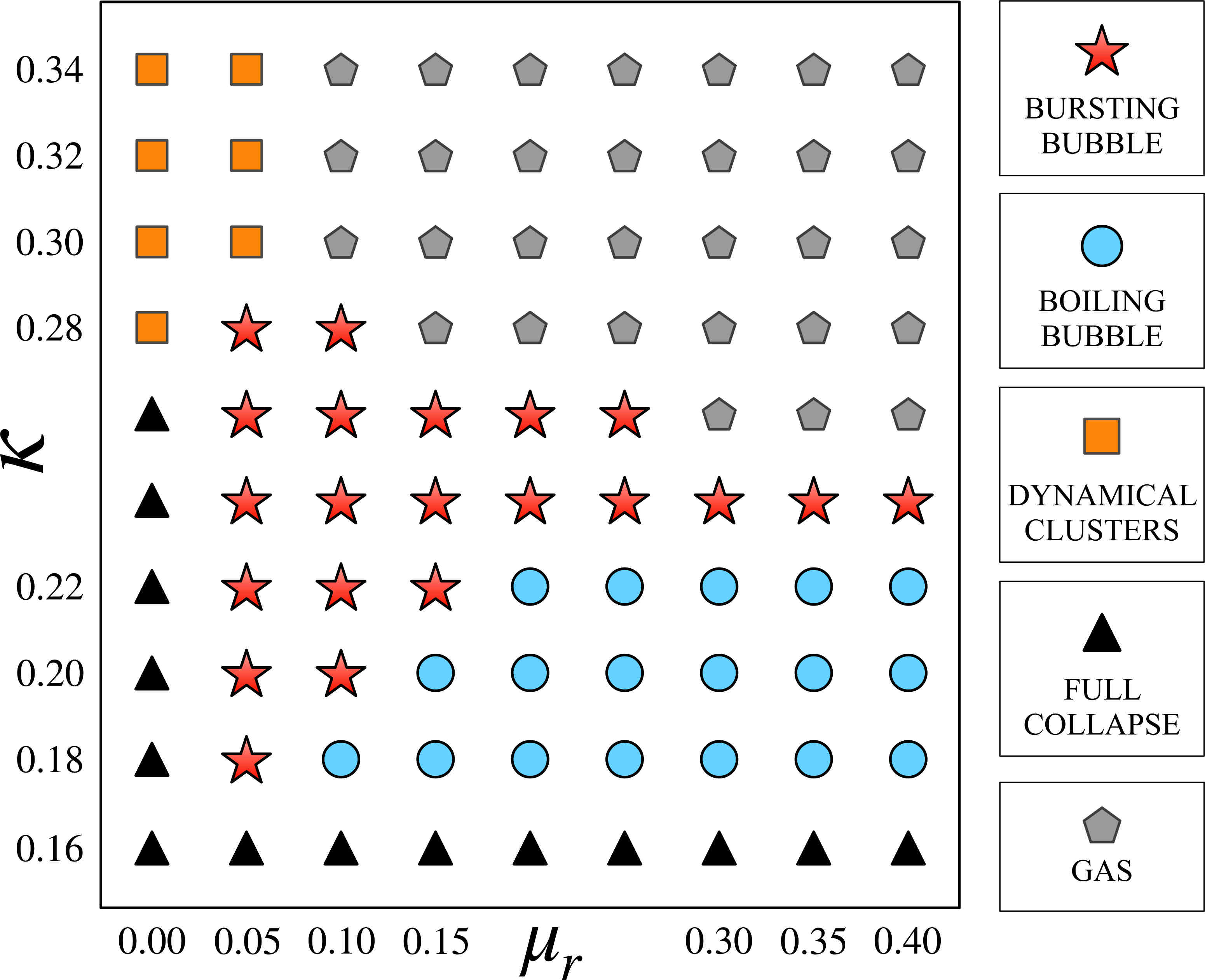}
 \caption{ Phase diagram in the plane of screening length $\kappa$ (see Eq. \ref{eq:current}) and strength of chemo-repulsive torque $\mu_r$ for the case of $\Lambda^c = 1$.  
    The five phases observed are: \textit{bursting bubbles}, \textit{boiling bubbles}, \textit{dynamical clusters}, \textit{full collapse} to a single cluster, and a \textit{gaseous} phase. The bubbly phases are obtained as intermediate phases between well known phases of full collapse and dynamic clusters.}
    \label{fig:phaseDiagramMarker}
\end{figure}

Here, two particles are considered bound if separated by less than $2.1b$, and a cluster consists of a bound set of more than 10 particles. To identify phases (illustrated in Fig.-\ref{fig:phaseDiagramMarker}), we use triad of metrics: the standard deviation of $S$ over the system, denoted as $\sigma_S$, and the coordination number $Q$, defined as the average number of neighbors per particle within the same cutoff distance $2.1b$. To construct the phase diagrams, all three quantities ($S$, $\sigma_S$, and $Q$) are subsequently time-averaged over the steady state.

\begin{figure*}
\centering
\includegraphics[width=\textwidth]{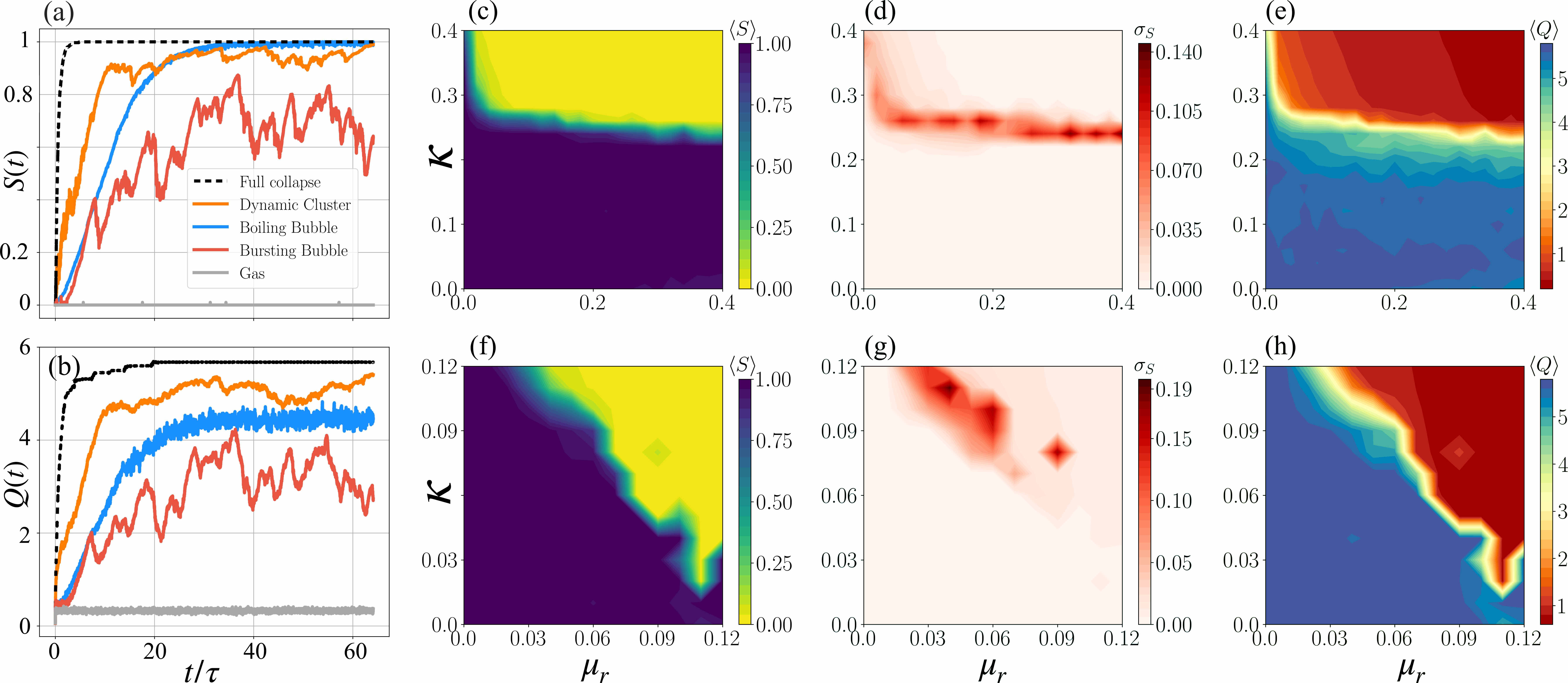}
\caption{Structural metrics of the system. (a) Time evolution of the fraction of total cluster size, $S(t)$. (b) Time evolution of the coordination number, $Q(t)$. The distinct phases — \textit{full collapse}, \textit{dynamical clusters}, \textit{boiling bubble}, \textit{bursting bubble} and \textit{gas} — are distinguished by their steady-state time averages ($\langle S \rangle, \langle Q \rangle$) and temporal fluctuation magnitude ($\sigma_{\scriptscriptstyle S}$).
Panels (c-h) are
phase diagrams of the system in the $(\mu_r, \kappa)$ parameter space for $\Lambda^c = 1$ (top row) and $\Lambda^c = 0$ (bottom row). The left column (c, d) shows the time-averaged cluster size fraction $\langle S \rangle$, the middle column (d, g) shows the temporal fluctuation of $S$ over time, $\sigma_S$, and the right column (e, h) shows the time-averaged coordination number $\langle Q \rangle$. Color maps indicate metric values used to map the five structural phase regimes.
}
\label{fig:Phase_diagram}
\end{figure*}

To uniquely classify the structural phases present in the system, we analyze the time evolution of the normalized cluster size, $S(t)$ (Fig. \ref{fig:Phase_diagram}a), and the coordination number, $Q(t)$ (Fig. \ref{fig:Phase_diagram}b). Phase distinction begins with the cluster size dynamics in Fig. \ref{fig:Phase_diagram}(a). The \textit{gas} phase is immediately evident from its near-zero cluster fraction ($\langle S \rangle \approx 0$). Though both of \textit{bursting bubble} and \textit{dynamical clusters} shows temporal fluctuations, $\sigma_{\scriptscriptstyle S}$ (Fig. \ref{fig:Phase_diagram}d,g), but \textit{bursting bubble}'s steady-state time average, $\langle S \rangle$ remains consistently lower than that of the \textit{dynamical clusters} ($\langle S\rangle \lessapprox 1$ for \textit{dynamical clusters}). Conversely, both the \textit{full collapse} and \textit{boiling bubble} phases display maximal coverage ($\langle S \rangle \approx 1$) with zero temporal fluctuation ($\sigma_S \approx 0$), making them indistinguishable from $S(t)$ alone. To separate these two condensed phases, we evaluate the time-averaged coordination number, $\langle Q \rangle$, from Fig. \ref{fig:Phase_diagram}(b). The \textit{full collapse} phase retains high structural packing characterized by $\langle Q \rangle > 5.6$, whereas the \textit{boiling bubble} phase shows reduced coordination with $\langle Q \rangle < 5.0$.

\begin{figure*}[t]
\centering
\includegraphics[width=0.96\textwidth]{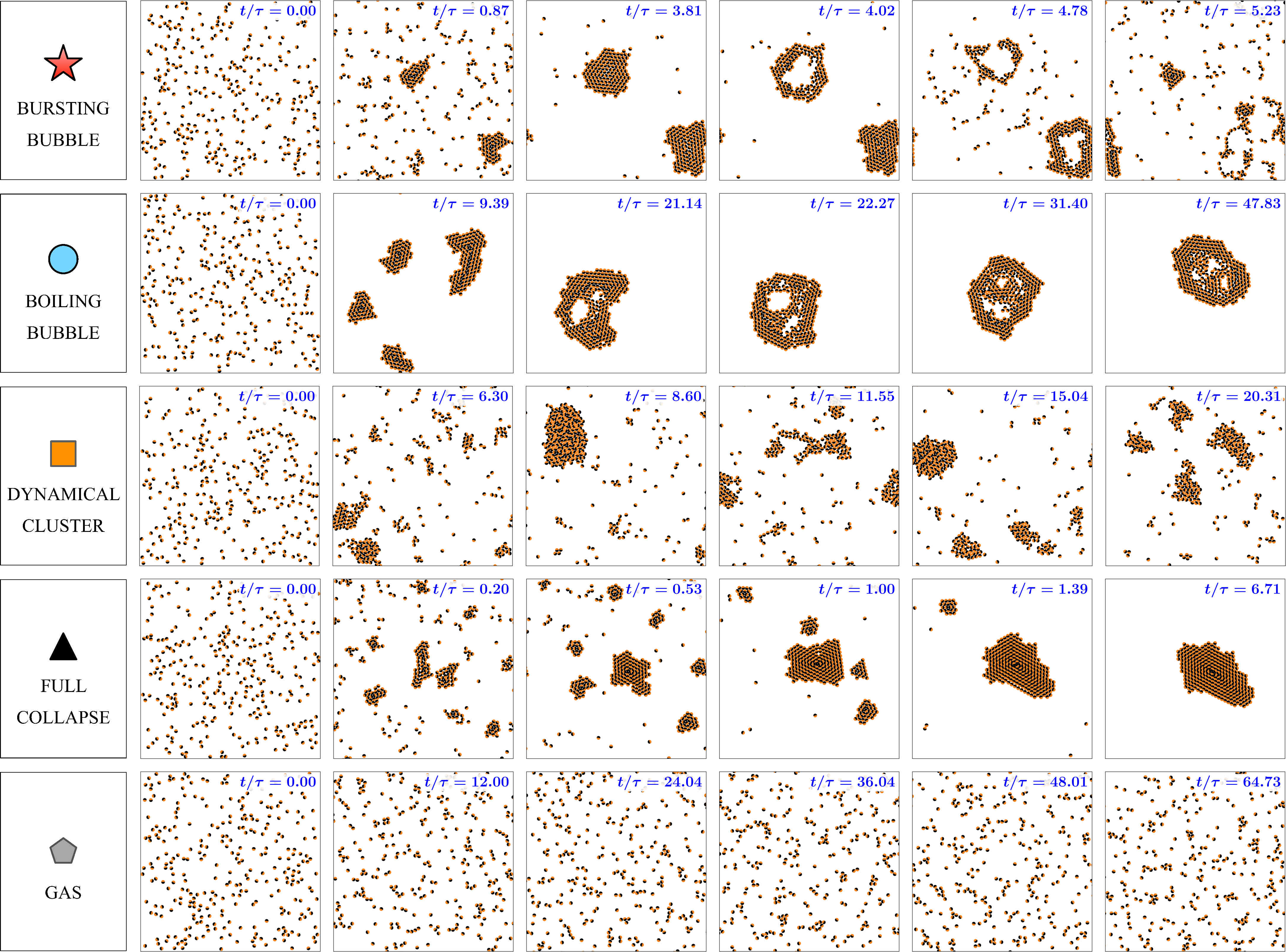}
\caption{The snapshots of evolution and formation of each phase, as illustrated in Fig.-\ref{fig:phaseDiagramMarker}, at different time steps. In every phase, we start with randomly distributed particle positions and orientations in the system. The first row depicts a \textit{bursting bubble}, where particles form small clusters that burst before transitioning into a fully collapsed cluster phase. The second row illustrates the behavior of a boiling bubble, demonstrating a manifestly non-equilibrium steady state distinct from a \textit{bursting bubble}. The third and fourth rows present the \textit{dynamical clusters} phase and the \textit{full collapse} phase, respectively. The last row shows the \textit{gas} phase, which we have observed for a long duration to demonstrate that it remains in this state indefinitely. Note that smaller systems of $N=300$ particles are shown in this figure for visual clarity, whereas all quantitative analyses throughout the manuscript employ $N=1000$ particles.}
\label{fig:phaseTime}
\end{figure*}

Phase boundary identification across the parameter space $(\mu_r, \kappa)$ is illustrated in (Figs. \ref{fig:Phase_diagram}c-e).
In the average cluster size $S$ maps (Figs. \ref{fig:Phase_diagram}c, d), low values of $\langle S \rangle \approx 0$ delineate the \textit{gas} phase. Elevated temporal fluctuations in $\sigma_S$ (Figs. \ref{fig:Phase_diagram}d, g) highlight candidate regions for either the \textit{bursting bubble} or \textit{dynamical clusters} regimes. For $\Lambda^c = 1$, the high-$\sigma_S$ domain near $\mu_r = 0$ coincides with maximal spatial coverage ($\langle S \rangle \approx 1$ in Fig. \ref{fig:Phase_diagram}c), uniquely identifying the dynamical clusters phase. 
The remaining high-$\sigma_S$ region, associated with intermediate $\langle S \rangle$, corresponds to the \textit{bursting bubble} state. 
See time evolution of these distinct phases in Fig.\eqref{fig:phaseTime}.

\subsection{Mechanism of Phases}
The emergent phase behavior in our system is governed by a delicate interplay among self-propulsion, long-range phoretic attraction, torque-induced particle reorientation, and rotational Gaussian noise. We set the translational mobility to $\mu_t = -1$ and consider positive rotational mobility ($\mu_r > 0$). The negative translational mobility ($\mu_t = -1$) ensures that a particle's self-propulsion points in the direction of its phoretic cap, while the phoretic interaction velocity consistently draws particles toward one another, resulting in persistent long-range attraction. Conversely, positive rotational mobility ($\mu_r > 0$) introduces chemical torques that tend to reorient neighboring particles to face away from one another. However, because translational velocity includes an explicit self-propulsion term ($\bm{V}_i^{\text{self}}$) alongside phoretic attraction, spatial aggregation occurs on a faster timescale than torque-driven particle reorientation. As a result, particles pull together into dense clusters before chemical torques have sufficient time to align neighboring particles in opposite directions (Movie S1).

This dynamic gives rise to a distinct core-shell architecture within assembled clusters. Particles located at the cluster boundary experience a strongly asymmetric chemical gradient, generating a robust outward-directed torque. Conversely, interior core particles are surrounded by an isotropic chemical flux from all sides, causing their net phoretic torque to vanish. With net chemical torques eliminated in the interior, rotational fluctuations can readily reorient core particles against the weak local field, whereas boundary particles remain locked facing outward by the strong chemical gradient.

This architectural asymmetry dictates the onset of the \textit{bursting bubble} phase. When stochastic rotational fluctuations cause a localized domain of interior core particles to spontaneously align, their combined self-propulsion exerts persistent pressure against the outer boundary shell. Because edge particles are locked outward, this internal mechanical stress disrupts the structural balance of the cluster, leading to explosive rupture. Furthermore, the survival of this mechanism depends sensitively on the competition among key parameters: if rotational noise is too intense (low $\mathrm{Pe}$), core orientations randomize too rapidly for coherent directional domains to form, suppressing cluster bursting. Similarly, at very low values of the screening parameter $\kappa$, strong phoretic attraction stabilizes the cluster against internal disruption, preventing bursting altogether. To systematically explore these phase regimes, we fix $\mathrm{Pe} = 100$ throughout this work.

\begin{figure}[t]
    \centering
    \includegraphics[width=0.8\linewidth]{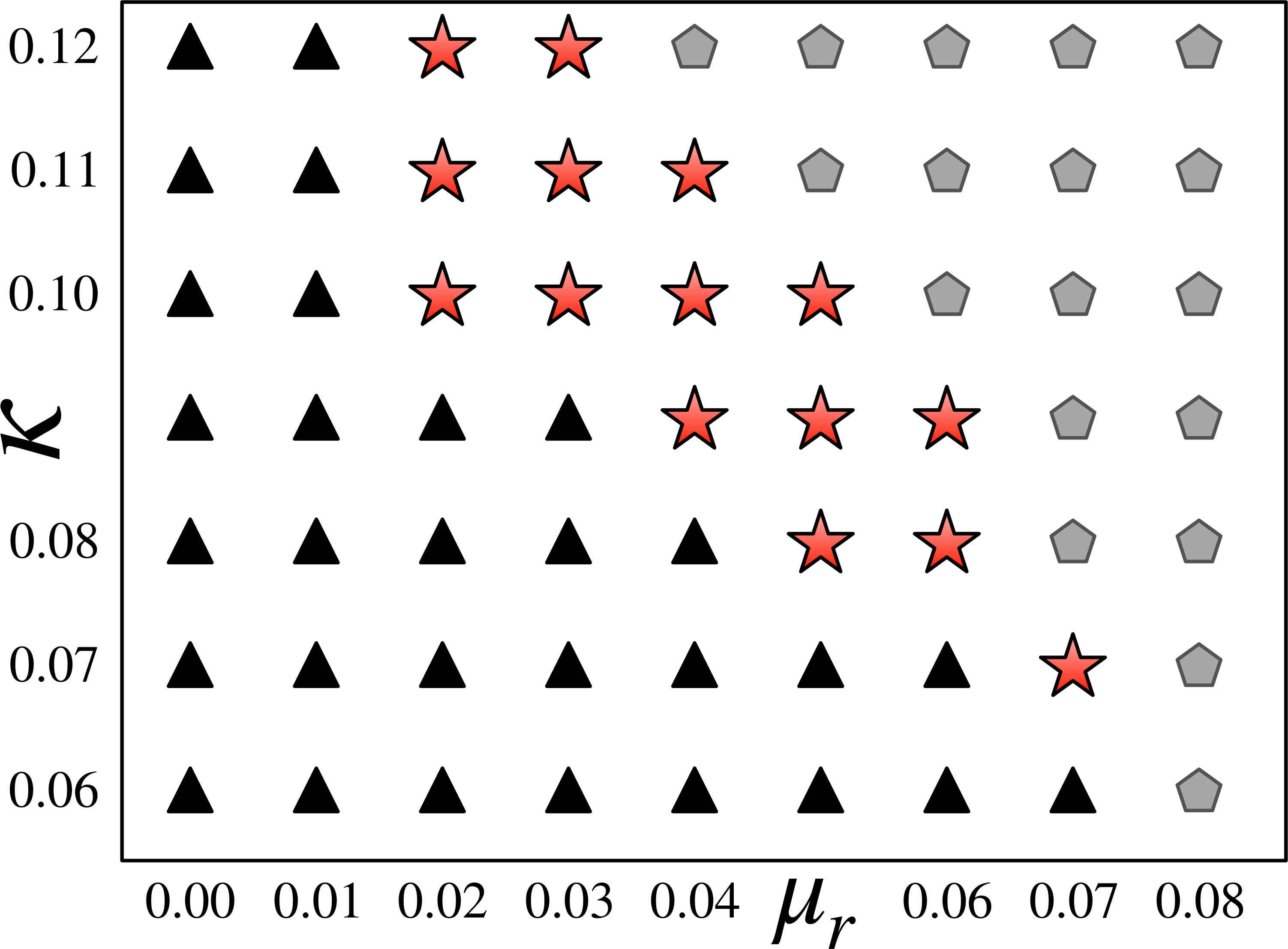}
    \caption{Phase diagram in the plane of screening length $\kappa$ (see Eq. \ref{eq:current}) and strength of chemo-repulsive torque $\mu_r$ for $\Lambda^c = 0$.
    Markers have same meaning as in Fig.\ref{fig:phaseDiagramMarker}.
    This figures show that we get bursting bubbles at $\Lambda^c=0$ in a very small parameter region. On the other hand, we have not seen any boiling bubble in the case of $\Lambda^c=0$. 
    }
    \label{fig:phaseDiagramMarker2}
\end{figure}

Increasing the rotational mobility $\mu_r$ alters the nature of these cluster disruptions depending on the screening parameter $\kappa$. At moderate screening values ($\kappa \approx 0.18\text{--}0.22$), increasing $\mu_r$ generates the \textit{boiling bubble} regime (Fig. \ref{fig:phaseDiagramMarker}a). As $\mu_r$ increases, chemical torques strengthen across the cluster; consequently, rotational fluctuations at fixed $\mathrm{Pe}$ can no longer align a sufficiently large domain of core particles in a single direction to overcome the cohesive outer ring. Instead of experiencing explosive rupture, localized core fluctuations cause particles to continuously circulate and push against the tightly locked perimeter shell, producing the \textit{boiling bubble} state. While multiple distinct boiling bubbles initially emerge in the system, they undergo coarsening over long timescales: whenever two boiling bubbles touch, they irreversibly coalesce into a larger cluster, eventually resulting in a single system-spanning boiling bubble at late times (Movie S2). However, at higher screening values ($\kappa \approx 0.24\text{--}0.28$), reduced phoretic attraction renders the outer ring unable to contain escaping particles, allowing core particles to break through the boundary unimpeded and driving a direct transition from the \textit{bursting bubble} state into the \textit{gas} phase. Notably, the nucleation of both \textit{bursting} and \textit{boiling} bubbles requires a critical cluster size; in very small systems or small clusters, central particles remain strongly confined and do not experience sufficient torque cancellation to form a distinct core domain, preventing bubble nucleation altogether.

In the absence of reorientation torques ($\mu_r = 0$) due to chemorepulsion, particles exhibit no intrinsic directional preference within clusters. Structural organization is then controlled entirely by the screening parameter $\kappa$, where strong attraction (small $\kappa$) maintains a \textit{full collapse} state, while reduced attraction (larger $\kappa$) leads to the \textit{dynamical clusters} regime. Interestingly, within the \textit{full collapse} regime, slightly increasing $\kappa$ from zero causes the cluster morphology to transition from circular to elongated (see Movie-II), as the weakened phoretic attraction can no longer pull all particles inward with sufficient force to sustain an isotropic, rounded cluster. A rigorous quantitative characterization of this cluster shape transition is beyond the current scope and remains a subject for future work.

Finally, at $\Lambda^c = 0$, the baseline magnitudes of phoretic attraction and torque are reduced, shifting the \textit{bursting bubble} phase to smaller values of $\kappa$ and $\mu_r$. 
See Figure \ref{fig:phaseDiagramMarker2}.
Due to this weakened attraction, a smaller $\kappa$—corresponding to a longer interaction range—is required compared to $\Lambda^c = 1$ to maintain cluster cohesion long enough for internal self-propulsion stress to build up and drive explosive bursting as $\mu_r$ increases.
Consequently, this domain consists entirely of bursting bubbles. In this parameter space at $\Lambda^c = 0$, the dynamical clusters phase is predominantly suppressed, appearing only within a narrow window at $\mu_r = 0$ within a narrow intermediate window ($\kappa \approx 0.20\text{--}0.24$), before terminating prior to $\kappa = 1.0$. Separating the dense static regimes—\textit{full collapse} and \textit{boiling bubble}—requires evaluating the time-averaged coordination number $\langle Q \rangle$ (Figs. \ref{fig:Phase_diagram}c, f). At $\Lambda^c = 0$, the \textit{boiling bubble} phase is not explicitly observed within the current parameter resolution. Because the \textit{boiling bubble} typically manifests as a narrow intermediate phase bordering the \textit{full collapse} regime, fine-grained resolution mapping around these phase boundaries may uncover its presence, forming a target for future investigation. A comparison of dynamics at three values of $\Lambda^c$ is shown in Fig.\ref{fig:comparision} of appendix.

\section{Summary and discussions}
\label{sec:conclusion}

In this paper, we have studied the role of boundary conditions in controlling the emergent dynamics in active suspensions. 
We have demonstrated that spatial confinement and chemical screening fundamentally reshape the collective phase behavior of autophoretic particles. By focusing on the parameter space where phoretic translational forces are chemo-attractive ($\mu_t < 0$) while rotational phoretic torques are chemo-repulsive ($\mu_r > 0$), we mapped out the non-equilibrium phase diagram controlled by the screening parameter $\kappa$, torque magnitude $\mu_r$, and boundary reflection coefficient $\Lambda^c$. Beyond the well-known regimes of chemotactic macro-phase separation (\textit{full collapse}) and active dynamic clustering, our central finding is the emergence of novel boundary- and screening-induced \textit{bubbly phases} -- specifically classified into \textit{boiling bubbles} and \textit{bursting bubbles}, two manifestly non-equilibrium steady states.

To rigorously distinguish these emergent states, we introduced a metric triad comprising the steady-state cluster fraction $\langle S \rangle$, its temporal fluctuation magnitude $\sigma_S$, and the mean particle coordination number $\langle Q \rangle$:
\begin{itemize}
    \item \textbf{\textit{Full collapse} Phase:} Characterized by dense, close-packed aggregates with maximal global coverage ($\langle S \rangle \approx 1$), high structural packing ($\langle Q \rangle > 5.6$), and negligible temporal variation ($\sigma_S \approx 0$).
    \item \textbf{\textit{Boiling bubble} Phase:} Retains high global cluster coverage ($\langle S \rangle \approx 1, \sigma_S \approx 0$), but displays significantly reduced local coordination ($\langle Q \rangle < 5.0$). This structural signature reflects the continuous nucleation, drift, and annihilation of localized fluid cavities (bubbles) maintained inside the dense active phase.
    \item \textbf{\textit{Bursting bubble} Phase:} Represents an intensely dynamic regime where finite-sized clusters undergo cyclical growth, void formation, and sudden structural rupture, resulting in intermediate time-averaged cluster fractions ($\langle S \rangle \lessapprox 1$) accompanied by prominent temporal fluctuations ($\sigma_S > 0$).
    \item \textbf{\textit{Dynamical clusters} Phase :} Dynamic clusters exhibit moderate temporal fluctuations ($\sigma_S > 0$) without cavity stabilization.
    \item \textbf{\textit{Gas} Phase:} The \textit{gas} phase displays a near-zero cluster fraction ($\langle S \rangle \approx 0$).
\end{itemize}

Crucially, our comparative study between a no-flux wall ($\Lambda^c = 1$) and a chemically permeable interface ($\Lambda^c = 0$) reveals that boundaries actively direct non-equilibrium phase separation. The presence of a no-flux boundary introduces image chemical sources that boost local concentration gradients, reinforcing reorientational torques near the cluster interface. This boundary-induced reinforcement stabilizes internal cavity formation, giving rise to robust \textit{boiling bubble} states that are completely absent when chemical screening occurs in an unconfined/permeable medium ($\Lambda^c = 0$).

Furthermore, tuning the inverse screening length $\kappa$ dictates the effective spatial reach of phoretic interactions. Increasing $\kappa$ systematically weakens long-range attractive forces, driving transitions from dense macroscopic collapse toward bubbly phases, dynamical micro-clusters, and ultimately a homogeneous \textit{gas} phase. Interestingly, increased screening also induces structural anisotropy in the dense phase, causing full clusters to transition from isotropic circular shapes to elongated geometries. See Fig.\ref{fig:aniso}.

Finally, we highlight important finite-size dynamics: a threshold particle number $N$ is essential to observe bubble nucleation. In smaller systems, particles at the center of a cluster remain strongly bound by cumulative attraction, preventing internal phoretic torques from forming a stable void. Over extended temporal scales, smaller localized bubbles within the bulk merge upon contact, eventually coalescing into a single macro-scale systemic bubble state.

Future work will focus on exploring intermediate and negative boundary reflection coefficients ($-1 \le \Lambda^c < 1$) to understand how chemical absorption at boundaries modifies active assembly. Additionally, a detailed morphological analysis of the screening-induced transition from circular to elongated clusters, alongside the incorporation of full hydrodynamics in three dimensions, will offer broader design rules for reconfigurable active metamaterials, and microfluidic transport systems.
In addition, exploring the competing roles of hydrodynamic and phoretic interactions and their modification by boundary conditions \cite{thutupalli2018flow} suggests directions for further work.
\section*{Acknowledgments.}
We thank Professor H.A. Stone for many helpful discussions. Numerical work was performed on the NSM PARAM RUDRA Supercomputing Facility PARAM Shakti at IIT Madras. We also thank the U.S. National Science Foundation for support of this research via the Princeton Center for Complex Materials, a MRSEC (NSF DMR-2011750).

\appendix 
\section{Derivation of EOM}\label{app:DeriveEOM}
We consider a system of Janus particles, half-coated with a catalytic material generating a flux of solutes. Latin indices ($i$, $j$, ...) are used for particle indices. Greek indices ($\alpha$, $\beta$, ...) are reserved for Cartesian coordinates. This yields the following boundary condition for the distribution of solute concentration $c_i$ on the surface $S_i$ of each particle,
\begin{equation}
    -D\bm{n}\bm{\cdot}\bm{\nabla}c_i(\bm{x})=
    \begin{cases}
        A_i,\quad \bm{x}\in\text{ catalytic cap,}\\
        0,\quad \,\;\bm{x}\in\text{ inert face.}
    \end{cases}
    \label{eq:activity}
\end{equation}
Here, $D$ is the local solute diffusivity in the suspension ($z>0$), which we identify with $D_1$ below. 
For simplicity, we assume that each particle is equally active, i.e. $A_i=A$.
In the limit of low P\'{e}clet number, the solutes diffuse in the bulk without being advected by the flow.

The slip velocity distribution on the surface of each particle is given by the phoretic boundary condition \citep{golestanian2007designing}
\begin{equation}
    \bm{v}^s_i(\bm{x})=\mu_i(\bm{x})\bm{\nabla}_s c_i(\bm{x})
    \quad\text{for}\quad \bm{x}\in S_i,
    \label{eq:phoretic-slip}
\end{equation} 
where $\bm{\nabla}_s c_i(\bm{x}) = (\mathbb{I}-\bm{n}\bm{n})\cdot\bm{\nabla} c_i(\bm{x})$ with $\bm{n}$ the unit normal vector to the surface of the particle.
\begin{align}
        \mu_i(\bm{x})=
    \begin{cases}
        \mu_{C,i},\quad \bm{x}\in\text{ catalytic cap,}\\
        \mu_{I,i},\quad\; \bm{x}\in\text{ inert face.}
    \end{cases}
\end{align}
For simplicity, we assume that all particles are identical Janus colloids with the same catalytic and inert face mobilities, $\mu_C$ and $\mu_I$. 

To the leading order, hydrodynamic interactions are sub-dominant compared to chemical interactions for motion of autophoretic particles in a plane parallel to the wall. 
Neglecting the former, we can summarise the linear and angular particle dynamics as follows \cite{anderson1989colloid, liebchen2017phoretic, liebchen2019interactions, turk2025autophoretic}: 
\begin{align}
    \bm{V}_i &= M_T\bm{F}_i 
            -\dfrac{1}{4\pi b}\int_{S_i}\bm{v}^s_i\,\mathrm{d}S,\\
    \bm{\Omega}_i&=-\dfrac{1}{8\pi b^3}\int_{S_i}\bm{n}\times\bm{v}^s_i\,\mathrm{d}S
            + \sqrt{2D_r}\,\bm{\xi}_i,
\end{align}
where $M_T=1/(6\pi\eta b)$ is the translational mobility of a sphere of radius $b$ in a fluid of viscosity $\eta$, and $\bm{F}_i$ denotes external forces such as the excluded-volume repulsion $\bm{F}^{\mathrm p}_i$ used in the simulations (Sec.~\ref{sec:model}).
Using the symmetry of half-coated spherical particles and neglecting terms of order $\mathcal{O}(\bm{\nabla}\bm{\nabla}c|_{\bm{r}_i})$ \cite{anderson1989colloid, liebchen2017phoretic}, the active contributions reduce to
\begin{align}
    \bm{V}_i &= M_T\bm{F}_i 
            - \tfrac{1}{4}\langle\mu\rangle\tfrac{A}{D}\,\bm{p}_i
            - \langle\mu\rangle\,\bm{\nabla}c\big|_{\bm{r}_i},\\
    \bm{\Omega}_i &= -\tfrac{9}{16}(\mu_C-\mu_I)\,\bm{p}_i\times\bm{\nabla}c\big|_{\bm{r}_i}
            + \sqrt{2D_r}\,\bm{\xi}_i,
\end{align}
where $\langle\mu\rangle=(\mu_C+\mu_I)/2$ and $\bm{p}_i$ points towards the catalytic cap.

To solve for the chemical field, we assume that the suspension is confined to the positive half-space $z>0$ by an infinite surface in the $x$-$y$ plane. At this surface, the solute obeys \citep{turk2025autophoretic}
\refstepcounter{equation}
\begin{align}
    c^{(2)} &= \kappa_c\, c^{(1)},\quad 
    D_1\,\partial_z c^{(1)} = D_2\,\partial_z c^{(2)}\;\,
    \text{ for }\;z=0\\
    c^{(i)}&\rightarrow 0\;\text{ for }\; r\rightarrow\infty,
    \label{eq:c-bc}    
\end{align}
where $c^{(i)}$ and $D_i$ with $i=1,2$ are the concentration field and solute diffusivity in the regions $z>0$ and $z<0$, respectively. The solute permeability $\kappa_c\in\{0,1\}$ indicates whether the surface is impermeable ($\kappa_c=0$) or permeable ($\kappa_c=1$) to the solutes. The particle is assumed to be the only source of solutes so that the solute concentration vanishes far from the particles.
We characterise the boundary through the reflection coefficient
\begin{equation}
    \Lambda^c = \frac{D_1 - \kappa_c D_2}{D_1 + \kappa_c D_2},
\end{equation}
where $\Lambda^c = 1$ represents a no-flux wall ($\kappa_c=0$), $\Lambda^c=-1$ represents a chemically attractive permeable fluid-gas boundary ($\kappa_c=1$, and $D_2/D_1\rightarrow\infty$), and $\Lambda^c=0$ represents a chemically permeable interface between equally diffusive media ($D_2=D_1$), i.e.\ the chemical field does not sense the boundary.

For simplicity, we further assume that the particles' orientations and dynamics are confined to a quasi-2D plane at a height $h$ (determined by a balance of gravitational, chemo-hydrodynamic, and other near-field effects) parallel to this underlying surface. 
The phoretic field is obtained by considering activity of each particle following Eq.\eqref{eq:activity}.
Henceforth, we non-dimensionalise lengths by the particle radius $b$, writing $\bm{r}_{ij}=(\bm{r}_i-\bm{r}_j)/b$ for the separation vector, $r_{ij}=|\bm{r}_{ij}|$ for its magnitude, and analogously $\bm{r}^*_{ij}=(\bm{r}_i-\bm{r}^*_j)/b$ with $\bm{r}^*_j=(x_j,y_j,-h)^T$. Gradients in what follows are taken with respect to these dimensionless coordinates.
We also assume a first-order sink $-\lambda_d c$ in the chemical field \cite{liebchen2019interactions}, so that in the bulk ($D\equiv D_1$)
\begin{equation}
    D_1\!\left(\nabla^2 - \frac{\kappa^2}{b^2}\right)c = 0,
\end{equation}
away from the point sources, where $\lambda_d = D_1\kappa^2/b^2$ and $\kappa = b\sqrt{\lambda_d/D_1} = b/\ell_s$ is the dimensionless inverse screening length, with $\ell_s=\sqrt{D_1/\lambda_d}$ the screening length.
At leading order, each particle $j$ acts as a monopole of strength $\langle A\rangle/(Db)$. Superposing the screened monopole and image solutions \cite{liebchen2019interactions,turk2025autophoretic} subject to Eqs.~\eqref{eq:c-bc} and~\eqref{eq:activity}, the chemical field at particle $i$ is
\begin{align}
c(\bm{r}_i)
&=\frac{\langle A \rangle}{D}\sum_{j=1}^N\first{\frac{e^{-\kappa r_{ij}}}{r_{ij}} + \frac{\Lambda^ce^{-\kappa r^*_{ij}}}{r^*_{ij}}}
\label{eq:chem-field}
\end{align}
where $r^*_{ij}=|\bm{r}^*_{ij}|$.
In the limit $\kappa\rightarrow 0$, Eq.~\eqref{eq:chem-field} reduces to the unscreened boundary value problem \citep{turk2025autophoretic}.
For $\Lambda^c=0$, the image term vanishes; in the additional limit $\kappa\rightarrow 0$, Eq.~\eqref{eq:chem-field} simplifies to the image-free chemical field used by \citet{pohlStarkPRL2014} and \citet{liebchen2019interactions} in similar analyses of the clustering behaviour of chemically interacting particles.
\begin{figure}[t]
\centering
\includegraphics[width=0.485\textwidth]{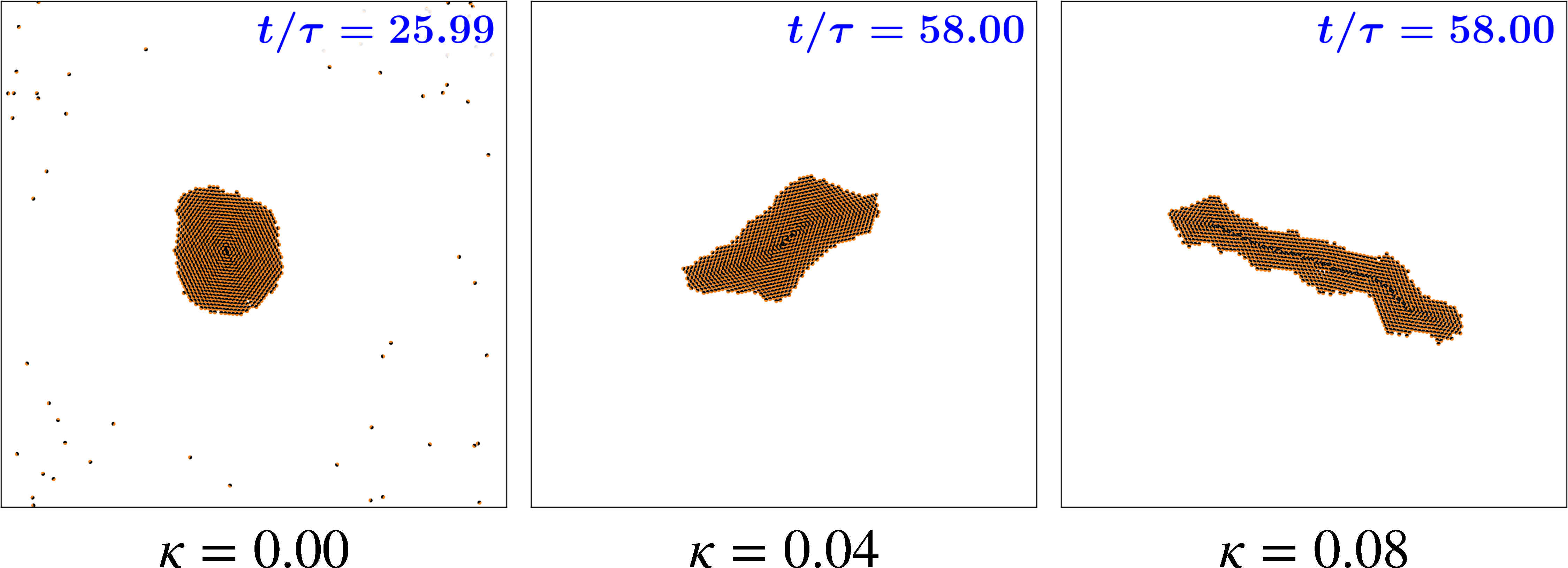}
\caption{Structural anisotropy of the macroscopic clusters at $\mu_r = 0.28$ for different values of screening $\kappa$ of phoretic interactions. On increasing $\kappa$, the clusters are more elongated. Here, $\Lambda^c=1$.}
\label{fig:aniso}
\end{figure}

We define the parameters 
\begin{align*}
v_0 &=\frac{|\langle \mu \rangle|A}{4D}\,,\qquad
\langle\mu\rangle=\frac{\mu_C + \mu_I}{2}
\\
\mu_t &= \frac{\mu_C + \mu_I}{|\mu_C + \mu_I|}\,, \qquad
\mu_r = \frac{\mu_C - \mu_I}{|\mu_C + \mu_I|}\,,
\end{align*}
characterising whether translational and rotational interactions are repulsive ($\mu_t,\mu_r>0$) or attractive ($\mu_t,\mu_r<0$), respectively. Note that $|\mu_t|=1$. For half-coated particles $\langle A\rangle = A/2$.
The dimensionless chemical current is
\begin{equation}
    \bm{J}_i = \frac{|\langle\mu\rangle|}{2v_0}\,\bm{\nabla}c\big|_{\bm{r}_i}.
\end{equation}
Differentiating Eq.~\eqref{eq:chem-field} and noting that $\bm{\nabla}(e^{-\kappa r_{ij}}/r_{ij}) = -(e^{-\kappa r_{ij}}/r_{ij}^2)(\kappa + 1/r_{ij})\,\bm{r}_{ij}$ (and analogously for the image term), we obtain Eq.~\eqref{eq:current}, for which  in the simulations, the sum is evaluated over $j\neq i$.
Substituting into the reduced dynamics above gives the self-propulsion and interaction terms in Eq.~\eqref{eq:partE}:
\begin{align*}
    -\langle\mu\rangle\,\bm{\nabla}c\big|_{\bm{r}_i} &= -\mu_t\,2v_0\,\bm{J}_i,\\
    -\tfrac{9}{16}(\mu_C-\mu_I)\,\bm{p}_i\times\bm{\nabla}c\big|_{\bm{r}_i} &= -\tfrac{9}{4}v_0\mu_r\left(\bm{p}_i\times\bm{J}_i\right),
\end{align*}
with self-propulsion speed $v_0=|\langle\mu\rangle|A/(4D)$ in the term $-\tfrac{1}{4}\langle\mu\rangle(A/D)\,\bm{p}_i=-\mu_t v_0\,\bm{p}_i$.
Together with the excluded-volume force $M_t\bm{F}^{\mathrm p}_i$ ($M_t=M_T$), this yields Eq.~\eqref{eq:mainLE}.
Terms arising from the particle interacting with its own image, or from the $j=i$ term in Eq.~\eqref{eq:current}, are $\mathcal{O}\left(h^{-3}\right)$ and shall henceforth be neglected.

\section{Simulation Methodology and Parameters}
\label{app:simDetails}
Particle positions and orientations evolve according to the dynamical equations in Eq. (1) using a forward Euler–Maruyama integration scheme, but with a strict upper limit of displacement, $ds = 0.01$ . The system employs periodic boundary conditions along both the $x$- and $y$-axes. Chemical interactions are evaluated using the minimum image convention with $(3\times 3)$ boxes.

\begin{figure*}
\centering
\includegraphics[width=0.9\textwidth]{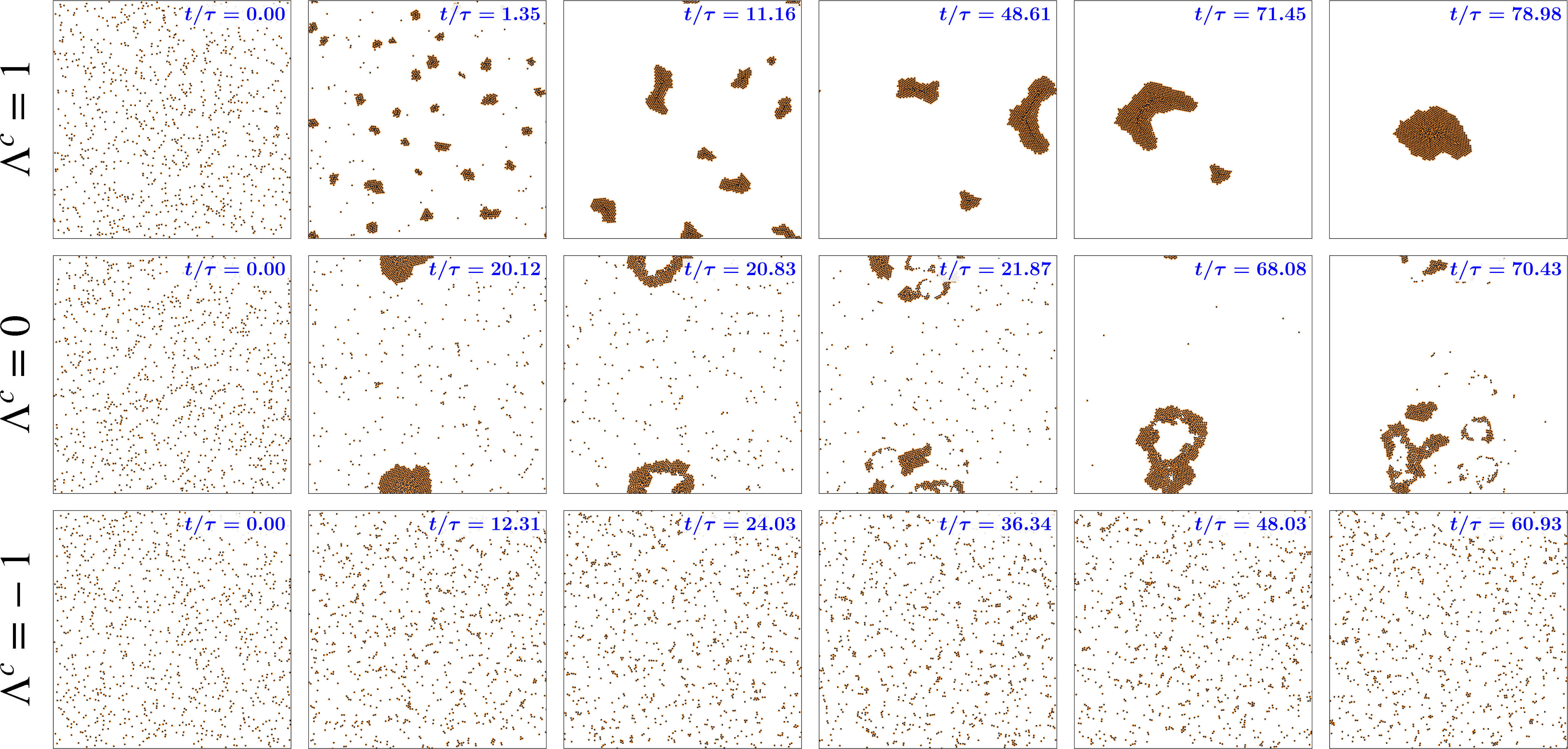}
\caption{Role of boundary conditions in steady-state behavior. Here, we only vary $\Lambda^c$, while keeping all other parameters fixed, such $\kappa = 0.08,\, \mu_r = 0.06$}
\label{fig:comparision}
\end{figure*}

The WCA potential $ U^{\mathrm{p}}$ 
to avoid overlap of particles is given as:
\begin{align}
    U^{\mathrm{p}}(r)=
    \begin{cases}
    \epsilon\left[
    \left(\dfrac{r_\mathrm{m}}{r}\right)^{12} - 2
    \left(\dfrac{r_\mathrm{m}}{r}\right)^6 + 1\right],
    \,\,\,\quad 
    \mathrm{when}\, r\leq r_\mathrm{m}
    \\
    0, 
    \qquad     \qquad     \qquad     \quad 
    \qquad     \qquad     \qquad
    \mathrm{otherwise}
    \end{cases}
\end{align}
 It is finite only when inter-particle separations are less than  $r<r_{\mathrm{m}}$ , where $r = ||\bm r_{ij}||$.
The constant $\epsilon$ sets the strength of the WCA potential.

\begin{table*}
\centering
\renewcommand{\arraystretch}{1.8} 
\setlength{\tabcolsep}{20pt}      
\begin{tabular}{| l | c | c |}
\hline
\multicolumn{1}{|c|}{\textbf{Parameter}} & \textbf{Sign} & \textbf{Value} \\ \hline
Radius of the particles   & $b$   & 1.0   \\ \hline
Height of the particles from the surface  & $h$   & 1.1   \\ \hline
Intrinsic velocity of the particles & $v_0$ & 2.0  \\ \hline
Number of particles & $N$ & 300, 1000, 10000  \\ \hline
Packing Fraction & $\Phi$ & 0.05  \\ \hline
Peclet Number & $\mathrm{Pe}$ & 100  \\ \hline
Strength of WCA potential & $\epsilon$ & 30  \\ \hline
\end{tabular}
\caption{Parameters used for simulations in this paper. We note that we have done simulations at different values of $N$. The phase diagram has been obtained by tuning $\mu_r$ and $\kappa$}
\end{table*}
\section{Supplemental Movies}
See supplemental movies at: \href{https://softmatter.gitlab.io/bubbles}{https://softmatter.gitlab.io/bubbles}. 
\section{Phase diagrams from simulations}
The phase diagram of Fig.\eqref{fig:phaseDiagramMarker} is made using more than one order parameter. This phase diagram is based on the 
one generated from simulations as shown in Fig.\eqref{fig:Phase_diagram}.

\end{document}